\documentclass[preprint,review, 12pt]{elsarticle}

\usepackage{amssymb}
\usepackage{amsmath}

\usepackage{graphicx}
\usepackage{amsmath}
\usepackage{amssymb}
\usepackage{lipsum}
\usepackage{multirow}
\usepackage{multicol}
\usepackage{makecell}
\usepackage{subfigure}
\usepackage[table]{xcolor}
\usepackage{float}
\usepackage{algorithmic}
\usepackage{algorithm}
\restylefloat{table}

\usepackage{booktabs}       
\usepackage{multirow}       
\usepackage{graphicx}       
\usepackage{caption}        
\usepackage{tabularx}

\usepackage{tcolorbox}
\usepackage{colortbl}

\renewcommand{\arraystretch}{1.5}

\definecolor{Blue}{gray}{0.85}
\newcolumntype{b}{>{\columncolor{Blue}}c}

\journal{Reliability Engineering \& System Safety}

\makeatletter
\def\ps@pprintTitle{%
  \let\@oddhead\@empty
  \let\@evenhead\@empty
  \def\@oddfoot{\reset@font\hfil\thepage\hfil}
  \let\@evenfoot\@oddfoot
}
\makeatother

\begin{document}

\begin{frontmatter}



\title{TARD: Test-time Domain Adaptation for Robust Fault Detection under Evolving Operating Conditions}



\affiliation[inst1]{organization={Intelligent Maintenance and Operations Systems, EPFL}, city={Lausanne}, country={Switzerland}}
\author[inst1]{Han Sun}
\author[inst1]{Olga Fink}

\begin{abstract}
Fault detection is essential in complex industrial systems to prevent failures and optimize performance by distinguishing abnormal from normal operating conditions. With the growing availability of condition monitoring data, data-driven approaches  have increasingly applied in detecting system faults. However, these methods typically require large, diverse, and representative training datasets that capture the full range of  operating scenarios, an assumption rarely met in practice, particularly in the early stages of deployment.
Industrial systems often operate under highly variable and evolving conditions, making it difficult to collect comprehensive training data. 
This variability results in a distribution shift between training and testing data, as future operating conditions may diverge from those previously observed ones. Such domain shifts hinder the generalization of traditional models, limiting their ability to transfer knowledge across time and system instances, ultimately leading to performance degradation in practical deployments.
To address these challenges, we propose a novel method  for continuous test-time domain adaptation, designed to support robust early-stage fault detection in the presence of domain shifts and limited representativeness of training data. Our proposed framework --Test-time domain Adaptation for Robust fault Detection (TARD) -- explicitly separates input features into system parameters and sensor measurements. It employs a dedicated domain adaptation module to adapt to each input type using different strategies, enabling more targeted and effective adaptation to evolving operating conditions. We validate our approach on two real-world case studies from multi-phase flow facilities, delivering substantial improvements in both fault detection accuracy and model robustness over existing domain adaptation methods under real-world variability.
\end{abstract}


\begin{highlights}

\item This work proposes TARD, a novel continuous test-time adaptation framework for fault detection under evolving operating conditions and limited data.

\item The framework is designed to address real-world industrial challenges by operating without any labeled fault data and requiring only a limited number of samples from the target system for adaptation.

\item To improve adaptation and reduce false alarms, the method separates system inputs into control variables and sensor measurements, which allows it to more effectively isolate faults from normal changes in operating conditions.

\item The proposed approach demonstrates substantial improvements and superior performance over existing domain adaptation methods in two real-world case studies involving multiphase flow facilities.
\end{highlights}


\end{frontmatter}

\section{Introduction}
\label{introduction}

Fault detection aims to identify evolving faults or degradation in complex industrial
systems to prevent failures, malfunctions, and unplanned downtime. Achieving early and robust fault detection is essential for optimizing system  performance, minimizing maintenance costs, and reducing asset unavailability.
Recent  advances in sensing technologies and the growing availability of condition monitoring data have facilitated the widespread adoption of  data-driven fault detection methods \cite{fink2020potential, xu2021machine}. 
However, these methods often rely on the assumption that large, diverse, representative training datasets are available -- covering all relevant operating and environmental conditions~\cite{hu2022prognostics}. In practice, such data are rarely available, particularly for newly installed or refurbished units that have limited historical data~\cite{rombach2021contrastive, ROMBACH2023108857}. 

To address this data scarcity,  one strategy is to leverage information from "experienced" fleet units -- systems with extensive operational histories and rich datasets. A viable solution is to directly apply models trained on these well-instrumented units to less-monitored or newly deployed ones. This cross-unit deployment aims to enhance fault detection performance across the fleet, particularly for units with limited training data. 
However, such model transfer is hindered by a fundamental limitation: data-driven models commonly assume that training and testing data are  independently and identically distributed (i.i.d)~\cite{nejjar2024domain}. In reality,  industrial systems  operate under diverse and dynamically changing conditions, violating this assumption. As a result,  significant distribution  shifts often exist between different units in the fleet. A model trained on one unit may therefore fail to generalize to another, frequently resulting in high  false alarm rates and reduced detection accuracy. These limitations ultimately undermine the ability to fully capitalize on shared fleet knowledge.

Domain adaptation (DA) has been extensively explored  to address distribution shifts in data-driven models \cite{michau2019domain, yan2024comprehensive}. These methods aim to align a labeled source domain with a related, unlabeled target domain. However, fault detection in industrial systems presents unique challenges.  First, DA approaches typically assume the availability of labeled data in the source domain, which is especially scarce or unavailable under faulty conditions in either the source domain or the target domain. As a result, many commonly applied unsupervised DA methods -- which rely on source domain labels -- are not applicable to fault detection tasks in industrial settings.
Second, most DA methods treat domain shift as a single, static distributional difference between discrete domains.
However,  in complex industrial systems, the operating conditions evolve continuously, leading to domain shifts even within the same unit over time. Therefore, domain adaptation should go beyond  discrete domain transfers and be capable of handling continuously evolving shifts. Relying on a fixed, discrete target domain is insufficient for ensuring robust fault detection  in such dynamic environments.  

To address these challenges, we propose Test-time domain Adaptation for Robust fault Detection (TARD), a novel framework for unsupervised, continual test-time domain adaptation for fault detection that operates entirely without labeled data in the source domain.  TARD enables robust fault detection across different units in industrial systems by continuously adapting to dynamically changing operating  conditions during test time. 
The key novelty of our approach lies in:
\begin{itemize}
    \item A continuous test-time adaptation strategy specifically designed for the unique challenges of unsupervised fault detection, where adapting to unlabeled test batches that potentially contain faulty samples.
    \item The explicit separation of input variables into control parameters and sensor measurements, with dedicated adaptation strategies for each. This allows the model to distinguish between changes in operating conditions (reflected in control variables) and system faults (reflected in sensor measurements), significantly reducing false alarms.
    \item A practical framework that requires only a limited number of normal samples from the target system, making it suitable for early-stage deployment and fleet-wide knowledge transfer.
\end{itemize}

TARD is based on a signal reconstruction framework for unsupervised fault detection.  Initially, a reconstruction network is trained to learn the system’s normal behavior,  allowing anomalies to be identified through deviations in the reconstruction residuals. To improve robustness under distribution shift, we incorporate a domain adaptation module that is trained on limited target samples and continually adapts to  incoming test batches. 
To distinguish between changes in operating conditions and actual faults, we partition  the input variables into two categories: exogenous parameters --  which may include both control settings and environmental conditions -- and sensor measurements. We then integrate separate domain adaptive strategies for each category, enabling the model to independently adapt to their respective distributions and more effectively isolate abnormal system behavior from normal variations.

By integrating continual test-time adaptation into the unsupervised fault detection paradigm, TARD enables fleet-wide knowledge transfer and real-time adaptation to evolving operating  conditions -- without the need for labeled data.
We validate TARD on two multi-phase flow facility cases studies, where it achieves significant performance gains over existing domain adaptation methods. The proposed framework is broadly applicable to other industrial systems, offering a scalable and robust solution for timely fault detection in dynamic and data-scarce environments.

\section{Related Work}

\subsection{Fault Detection} \label{fault detection}
Prognostics and Health Management (PHM) aims to improve asset reliability  and reduce operational costs by enabling 
accurate detection, diagnosis, and prediction of the remaining useful life (RUL). It encompasses three core functions: fault detection, which identifies abnormal system behavior using real-time monitoring data; fault diagnostics, which isolates the fault, determines its root cause, and classifies the fault type; and prognostics,  which predicts how long the system can continue operating before experiencing a failure or falling below acceptable performance levels \cite{fink2020potential}. Together, these capabilities support informed decision-making for maintenance planning and operational safety. 

Despite the promise of PHM techniques, applying them in real-world industrial settings  presents a major challenge: the scarcity of labeled faulty data~\cite{zio2022prognostics}. In critical systems such as hydropower plants, chemical processing facilities or railway systems, failures are rare by design, and degradation processes often  unfold gradually over long  periods. As a result, faults are either severely underrepresented or completely missing from most collected datasets, making supervised learning approaches difficult to apply effectively.

Given this limitation, the common approach is to first shift the objective to detecting any form of abnormal behavior in the system. This process begins with unsupervised anomaly detection, which involves learning a model of normal system operation and then identifying any data patterns that deviates from the baseline~\cite{venkateswara2017deep}. 
These detected anomalies indicate potential problems, which can be analyzed further to determine if they represent a fault that could lead to a failure of the system~\cite{wang2012survey, zio2022prognostics, nelson2024machine}.
  
These unsupervised learning methods can be broadly categorized into three main categories: probabilistic models, one-class classification methods, and reconstruction-based approaches~\cite{ruff2021unifying}. 
Probabilistic models aim to approximate the  probability distribution of normal data, using the estimated density  or an associated score function to assign  anomaly scores, where lower values typically indicate more anomalous behavior. Various probabilistic  models have been applied for this purpose, ranging from classic density estimation models (e.g. gaussian mixture models)~\cite{bishop1993novelty} to energy-based models (EBMs) \cite{zhai2016deep}, predicting anomalies through estimation of the normal data probability distribution.
One-class classification models take a different approach by  learning a discriminative decision boundary that encloses  the region of normal data, without explicitly estimating the full data distribution \cite{ruff2021unifying, Michau_2019, wang2023asymmetrical, he2024anomaly}. These models aim to learn a compact representation of normal behavior and flag data samples that significantly deviate from the normal data distribution as an anomaly\cite{wang2018unsupervised, zhang2021anomaly}.
Reconstruction-based methods, such as autoencoders (AEs), are trained to accurately reconstruct normal data samples. Anomalies are then identified based on  reconstruction errors --when a test input deviates significantly from the learned normal patterns, the model struggles to reconstruct it, resulting in a high error that signals a potential fault  \cite{lai2023context, hu2022low}. These models are expected to fit the data distribution under healthy conditions and then raise an alarm for predictions with large deviations when the test data distribution is significantly different from the learned distribution. Despite their usefulness, these unsupervised approaches require large, diverse, and representative datasets that comprehensively cover normal operating conditions, which is rarely fulfilled during the early operational phase of newly deployed systems, making robust early fault detection particularly challenging.


\subsection{Fleet Approaches for Fault Detection} \label{fleet}
As discussed above, unsupervised fault detection relies on the assumption that all possible normal conditions of the system can be learned from a sufficiently large and representative training dataset. 
This assumption is often violated in early operational phases -- such as for newly deployed or recently refurbished systems -- where data is limited and potentially biased toward transient startup behaviors. Under such scenarios, models trained on local data alone may fail in anomaly detection, especially under conditions not yet observed.

To alleviate the problem of limited early-stage data, fleet-level fault detection has gained interest as a promising approach. 
 A fleet can be defined from two perspectives: that of an operator or that of a manufacturer~\cite{michau2018fleet}. From the operator's perspective, a fleet consists of a group of assets that are owned and managed by the same organization for a specific operational purpose, and are typically used in similar ways and under similar conditions, resulting in relatively homogeneous operational profiles~\cite{jin2015comprehensive}. In contrast, from the manufacturer’s perspective, a fleet refers to a set of systems with similar design and functional characteristics, produced by the same manufacturer, but operated by different users in different environments~\cite{leone2017data}. In this work, we adopt the second broader definition - which encompasses more units and reflects greater variability in operating conditions -- an essential consideration for fault detection models.

 Fleet-based fault detection leverages operational data from the fleet for fault detection across systems. 
By leveraging data from a fleet of similar unit, models trained on data-rich systems can potentially improve fault detection and diagnostics of newer or less-instrumented units~\cite{leone2017data}. A good example would be a fleet of gas turbines or cars produced by the same manufacturer, while operating under varying conditions with different system configurations in different parts of the world \cite{fink2020potential}.

However, most of the existing approaches rely on the assumption that systems within the fleet behave similarly enough for models to transfer effectively~\cite{nejjar2024domain}. In reality, differences in environmental and operational conditions introduce significant variability.
This variability violates the assumption of independent and identically distributed (i.i.d.) data between units, often resulting in significant degradation in model performance when directly applied to different systems~\cite{zhao2024domain}. 

Early studies have attempted to address this by clustering  units that are similar enough to form homogeneous sub-fleets \cite{leone2016data, liu2018ai, azar2022semi}. To improve scalability and generalization, other approaches develop models that learn the overall fleet behavior and then identify similar units based on this learned functional representation~\cite{michau2018fleet}. While effective in some cases, such methods depend on the entire fleet sharing sufficient similarity, thus failing when units under homogeneous conditions do not exist or cannot be identified. 
More recent research has turned to domain adaptation to address distribution mismatch between units or between different operating conditions within the same unit ~\cite{michau2019unsupervised, yan2024comprehensive}, a topic discussed in section \ref{domain adaptation}.

\subsection{Domain Adaptation Applied to Fault Detection} \label{domain adaptation}
 
Domain adaptation (DA), a subfield of transfer learning, aims to align the distributions of source and target domains to mitigate domain shift~\cite{pan2009survey, wang2018deep}. DA  has been widely applied in PHM,  with both discrepancy-based methods \cite{zhang2022class, qian2023maximum} and adversarial-based methods \cite{michau2021unsupervised, qian2023deep, nejjar2024domain}, demonstrating promising performance for cross-domain scenarios where operational conditions  differ between training and deployment environments. 

However, most existing DA techniques rely on strong assumptions. They typically presume  that abundant source and target samples are available to capture the underlying data distributions. 
Mean discrepancy-based methods~\cite{qian2023maximum} require concurrent access to both abundant source and target domain data for distribution alignment, along with labeled samples for training regularization. Adversarial methods such as Domain Adversarial Neural Networks (DANN)~\cite{ganin2016domain} adapt to target distributions through domain discrimination between source and target domains, yet simultaneously demand labeled source data as training supervision.
This assumption rarely holds in industrial settings, particularly for newly installed systems where only limited data can be collected, impeding timely condition monitoring \cite{michau2021unsupervised}. Access to source data may also be restricted due to privacy or proprietary concerns, making unsupervised domain adaptation unfeasible \cite{fink2020potential}.

Methods such as the one proposed in~\cite{michau2021unsupervised} adopt a one-class classification approach to address the challenge of limited data samples.
However, they are generally designed for adaptation to one or more discrete, static target domains. In reality, industrial systems often operate under continuously evolving conditions due to environmental fluctuations, component aging, and ongoing system upgrades, leading to non-stationary distributions that static DA models cannot adequately address \cite{wang2022continual}. 
This limitation is particularly evident in applications such as energy systems and manufacturing processes, where operating conditions change continuously rather than shifting between discrete states.
As a result, domain adaptation strategies tailored  for discrete or static domains are ill-suited to dynamic industrial environments, where continuous, on-the-fly adaptation to domain shifts is required.

Recent advances in test-time adaptation (TTA) have emerged as a promising direction to address the limitations of conventional DA approaches.  TTA methods adapt pre-trained models during inference by leveraging only the distribution of the current test batch distribution~\cite{wang2022continual}. Notable TTA methods include Tent~\cite{wang2021tent}, which updates batch normalization statistics during inference, and SHOT~\cite{liang2020we}, which combines information maximization with self-supervised pseudo-labeling.
By dynamically updating model statistics or minimizing prediction entropy at inference, TTA enables real-time adaptation to distribution shifts, making it particularly attractive for PHM applications in environments with continuously evolving operating conditions \cite{wang2019domain}.

While TTA offers significant potential for adapting to dynamic changes, it has limitations in its application to fault detection. Although TTA requires only unlabeled target data at test time, most existing TTA approaches rely on labeled source data for pretraining the model. Additionally, they require continued access to labeled source data during adaptation as a regularization to prevent overfitting caused by aligning to the test distribution without supervision.
In the context of unsupervised fault detection, where fault are identified based on deviations from normal behavior, applying TTA to incoming  batches with unknown labels can inadvertently lead the model to adapt to, and thus incorporate, potentially faulty data within this batch as if it were normal. This may mask anomalies as in-distribution data, ultimately diminishing the model’s ability to detect  faults through predicted residuals. 

To conclude, achieving robust early-stage fault detection in PHM presents   several critical challenges. 
First, labeled fault data are often scarce or entirely unavailable in both source and target domains. Second, early-stage target data are typically  limited and may not be representative of the full range of the system's operating conditions. Third, industrial systems operate in continuously evolving environments, leading to  non-trivial distribution shifts that static domain adaptation methods struggle to address. 

Existing methods tend to address isolated aspects of these challenges, but rarely in combination.
Fleet-based approaches enable newly deployed  units to leverage fleet-wide knowledge, but often fail to adapt to the significant  variability of  operating conditions within the fleet itself. 
While DA methods have gained traction, most cannot handle continuous domain shifts and depend on access to sufficiently representative source and target data. Moreover, TTA in the context of unsupervised detection risks model overfitting due to the absence of supervised  faulty data. To the best of our knowledge, there is currently no approach that enables  unsupervised fault detection under dynamically changing, data-scarce environments.
In summary, prior research has made progress in addressing data scarcity and adaptation challenges for fault detection, but several important gaps remain. Our method is designed to fill these gaps through the following distinct contributions:
\begin{itemize}
    \item \textbf{Minimal Data Requirements:} Most conventional UDA approaches rely on substantial target domain data and often require labeled source data for regularization. However, in many industrial settings, only limited normal operational data is available from the target domain and faulty data is absent. Our method adapts effectively under these realistic constraints, requiring neither large datasets nor labeled faults.
    
    \item \textbf{Dynamic Operational Environments:} Existing methods typically assume static domain characteristics, limiting their robustness in real-world applications where operating conditions evolve over time. Our approach explicitly addresses this gap by continuously adapting to non-stationary distribution shifts in industrial systems.
    
    \item \textbf{Reliable Adaptation in Unsupervised Detection:} Existing test-time adaptation methods can incorrectly incorporate fault patterns as part of normal system behavior in unsupervised contexts. Our approach overcomes this limitation through the explicit separation and independent adaptation of control variables and sensor measurements, thereby enabling robust fault detection under  domain shifts.
\end{itemize}


\section{Preliminaries}

\subsection{Problem Definition}

The primary motivation for this research is to enable the transfer of operating experience across systems within a fleet -- specifically, from systems with abundant monitoring data to other systems operating while enabling adaptation ability to evolving operating conditions. A key challenge addressed in this work is the complete absence of labeled  fault data  in both  the source and target domains. We assume that only healthy operating data  is available  and that neither  the source nor target systems have experienced any faults during the training period. 
Additionally, we focus on target systems that  are newly taken into operation, where  only a limited number  of observations can be collected for training. Moreover, the data distribution for these systems can evolve continuously due to changes in operating conditions and environmental factors. The objective is to adapt a prediction model trained on the original system so that it can provide accurate predictions for new systems and fleets, even when only a small amount of training data is available. Given:
\begin{itemize}
    \item abundant healthy training data from the source system: $X_s = \left[x_1^s, \cdots, x_n^s\right]$, where $s$ denotes the source domain and $n$ denotes the number of data samples from the source domain, and
    \item limited observed normal data from the target domain: $X_t = \left[x_1^t, \cdots, x_m^t\right]$, where $t$ denotes the target domain and $m$ denotes the number of available data samples from the target domain,
\end{itemize}
the goal here is to achieve robust fault detection in the target domain $t$.

The proposed method takes into account limited data availability and varying operating conditions, specifically addressing scenarios with:
\begin{itemize}
    \item the absence of faulty training data in both domains,  
\item limited availability of target domain data for adaptation, and  
\item continuous distribution shifts that occur during inference as the system operates under changing conditions.
\end{itemize}

\subsection{System Variables} \label{system variables}
In industrial systems, input data typically comprise a mixture of variables reflecting different aspects of system behavior. To effectively handle domain shifts while preserving anomaly detection performance, it is important to distinguish between variables that represent changes in operating conditions and those that reflect the operational state of the system. Therefore, we define two groups within the input parameters: $X = [x, w]$.
$w$ denotes control variables, indicating variables that control system conditions. These variables are set by the operators or by the control system to optimize the performance under specified conditions.
$x$ represents sensor measurements, which are sensor signals monitoring system components and reflecting real-time system states. 
Here, we assume that changes in the distribution of control variables do not necessarily indicate an abnormal status but rather distinct operating conditions to which we should adapt.

\subsection{Reconstruction-based Anomaly Detection Framework} \label{reconstruction framework}
\begin{figure}[ht]
    \centering
    \includegraphics[width=0.95\linewidth]{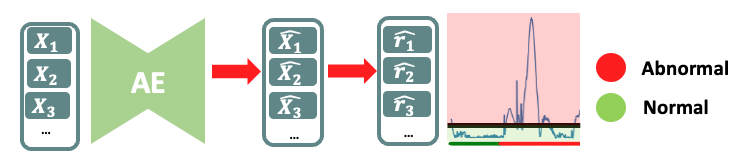}
    \caption{General pipeline of reconstruction-based unsupervised anomaly detection}
    \label{fig:pipeline}
\end{figure}

We deploy the paradigm of reconstruction-based pipeline for unsupervised anomaly detection as depicted in Figure \ref{fig:pipeline}. This method assumes that a model trained exclusively on normal (healthy) data can learn to reconstruct typical system behavior and deviations from this behavior with large reconstruction errors can be interpreted as potential anomalies.
We deploy an autoencoder (AE), denoted as $f_\theta$, as our reconstruction model. It is trained exclusively on normal source data samples $X_s$ to accurately model the normal data distribution with accurately predicted signal values $\hat{X_s}$. The training objective is to minimize the mean-squared error (MSE) between the original data samples $X_s$, and their reconstructed counterparts  $\hat{X_s}$:
\begin{equation}
    loss_{MSE}=\frac{1}{n} \sum_{1}^{n}\left(X_s-\hat{X}_s\right)^2
\end{equation}
where $n$ denotes the number of training samples.
Thus, on the healthy source dataset, we expect a small residual value $\hat{r_s} = \hat{X_s} - X_s$.
During testing, data samples generating large residuals are considered out-of-distribution and subsequently labeled as anomalies.

\section{Methodology}
\label{methodology}
\begin{figure*}[ht]
    \centering
    \includegraphics[width=1.0\linewidth]{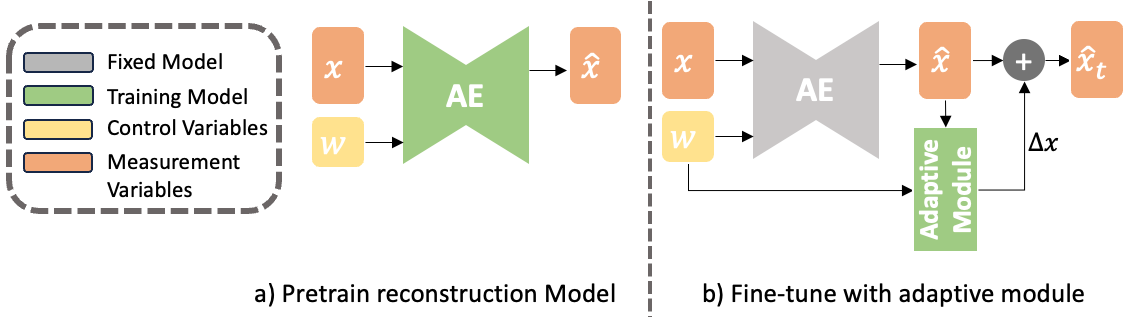}
    \caption{Test-time domain Adaptation for Robust fault Detection (TARD). a) We first pretrain the reconstruction-based anomaly detection model on the source dataset; b) For domain adaptative prediction, we add a domain adaptive module and train it on the target training data.}
    \label{fig:framework}
\end{figure*}

\subsection{Overview of TARD}
To address the challenge of robust fault detection, we propose a novel framework: Test-time domain Adaptation for Robust fault Detection (TARD).
TARD addresses three key challenges simultaneously: the absence of labeled fault data, limited data in the target domain, and continual adaptation to evolving operating conditions at inference time.

TARD consists of three core components: (1) a reconstruction-based anomaly detection model, (2) a test-time adaptation module, and (3) a scoring and decision mechanism.


Figure \ref{fig:framework} provides a high-level illustration of the TARD framework. TARD builds upon a standard reconstruction-based anomaly detection pipeline. It first trains an autoencoder that learns the normal operational patterns from abundant healthy data in a source domain. 
To adapt to novel distributions, rather than modifying the reconstruction model directly, we introduce a test-time adaptation module which compensates for shifts in operating conditions by predicting corrections to the autoencoder’s output. A residual-based scoring is smoothed over a temporal window and compared against a statistical threshold to identify faults. 

\subsection{Test-time Domain Adaptation Anomaly Detection}
The core of the TARD framework is its ability to adapt to changing operating conditions during inference. As discussed in Section \ref{system variables}, control variables 
$w$ represent operational inputs that are manually or automatically adjusted during system operation. These variables indicate intentional changes in the system's configuration and are not themselves indicators of abnormal behavior. Sensor measurements $x$, in contrast, reflect the real-time state of the system and are the primary source for fault detection. Therefore, as shown in Figure~\ref{fig:framework}(a), for the reconstruction-based anomaly detection model, control variables are included alongside sensor measurements as part of the input to the autoencoder while the reconstruction target consists only of the sensor measurements $x$. By excluding control variables from the reconstruction target and subsequent anomaly scoring, the model reduces the likelihood of false alarms triggered by expected changes in operating conditions.

Based on the modified reconstruction framework introduced, we integrate an additional adaptive module, $h_\phi$, for test-time domain adaptation to bridge the domain gap between the source and target domains. 
The decision to incorporate a separate adaptive module, $h_\phi$, rather than embedding adaptive layers directly into the reconstruction model, stems from limitations observed in unsupervised anomaly detection. 
TTA methods, such as AdaBN, adapt to each batch during test time, inevitably fitting the distribution of abnormal data points.  This impairs the model's ability to distinguish between normal and abnormal samples. Instead, our adaptive module takes the predicted value and original controlled system variables as inputs. By excluding monitoring data signals from adaptation, this approach prevents overfitting to the potentially faulty data distribution, thereby preserving the model's capability to accurately distinguish anomalies.

During the adaptation phase, the pre-trained autoencoder $f_\theta$ is frozen, and the adaptive module $h_\phi$ is trained on a few target data samples to predict $\Delta x$, aimed at compensating for the large prediction errors due to the domain gap between source and target data. This adaptive module exclusively processes the control variables $w$ as its input. To address continuous domain shifts during test time, an AdaBN layer is incorporated into the adaptive module.  This layer updates its mean and variance based on batch statistics during test time. The predicted $\Delta x$ is then added to the original prediction made by $f_\theta$ to compensate for inaccurate predictions caused by operating condition domain shift.
The training procedure for the adaptive module is detailed in Algorithm 1.

\begin{algorithm}[H]
\caption{Fine-tune Adaptive Module on Target Domain Data}
\begin{algorithmic}[1]
\REQUIRE Pretrained autoencoder $f_\theta$ on source domain data, target domain data $X_t$, number of epochs $E_{adapt}$, learning rate $\eta_{adapt}$
\STATE Initialize adaptive module $h_\phi$ with random weights
\STATE Freeze weights of $f_\theta$
\STATE Initialize optimizer (e.g., Adam) with learning rate $\eta_{adapt}$
\FOR{epoch $= 1$ to $E_{adapt}$}
    \FOR{each batch $B$ in $X_t$}
        \STATE Forward pass through autoencoder: $\hat{B} = f_\theta(B)$
        \STATE Forward pass through adaptive module: $\Delta x = h_\phi(B_{control})$
        \STATE Adjust reconstruction: $\hat{B}_{adjusted} = \hat{B} + \Delta x$
        \STATE Compute reconstruction loss: $\mathcal{L}_{MSE} = \frac{1}{|B|} \sum_{i=1}^{|B|} (B_i - \hat{B}_{adjusted,i})^2$
        \STATE Backward pass: Update $h_\phi$ using $\nabla \mathcal{L}_{MSE}$
    \ENDFOR
\ENDFOR
\STATE \RETURN Fine-tuned adaptive module $h_\phi$
\end{algorithmic}
\end{algorithm}

\subsection{Scoring and Fault Detection} \label{anomaly detection}
During test time, we compute the fault label $y\in[0, 1]$ based on the reconstruction result. $0$ denotes a healthy sample while $1$ indicates a faulty sample. Given the $i_{th}$ data sample $X_i = [x_i^1, .... x_i^k]$, we compute its relative residual:

\begin{equation}
\centering
    r_i = \frac{|\hat{X_i} - X_i|}{\bar{X}_{t\_training}}
\end{equation}

given its predicted reconstruction result $\hat{X_i}$. $k$ indicates the input dimension. $\bar{X}_{t\_training}$ represents the mean value of target data samples for training (including validation data), which helps scale the residual values. The anomaly score $s_i$ for each indidual data sample is calculated by integrating the scaled residual values across all of its $k$ sensor measurements:
\begin{equation}
    s_i = \frac{1}{k} \sum_{j=1}^{k}r_i^j
    + max\sum_{j=1}^{k}r_i^j
\end{equation}

To avoid false detection by outliers with extremely large residuals, the computed anomaly score is smoothed within a defined window length $l$ across the time dimension:
\begin{equation}
    s_{i\_smooth} = mean\sum_{q=0}^{l-1}s_{i+q}
\end{equation}

Faults are then detected based on $s_{i\_smooth}$, using a threshold determined via statistical analysis of the healthy validation set. We identify the data sample $X_i$ as an anomaly if:
\begin{equation}
    s_{i\_smooth} > \alpha * \bar{r}_{t\_training}
\end{equation}
where $\alpha$ is set empirically with a trade-off between the reduction of false alarms and sensitivity to faults.

\section{Case Studies}

We evaluate our proposed method on two multiphase flow facility datasets,  highlighting its effectiveness in achieving early fault detection and minimizing false positive alarm rates. We first present the implementation details common to both studies. Following this, we introduce each case study and the corresponding experimental results.

\subsection{Implementation details}

\noindent\textbf{Architecture} The autoencoder $f_\theta$ architecture consists of two parts: an encoder $f_e$ and a decoder $f_d$. $f_e$ comprises three fully-connected layers each followed by a batch normalization layer and a ReLU activation function, which consecutively map the original signal input to feature dimensions of 50, 50, and 10. $f_d$ follows a similar architecture but without batch normalization layers, decoding the latent representation from 10 to 50, 50, and then back to the original signal dimension. 

The adaptive module $h_\theta$ is a simple network composed of two fully connected layers that map the group of control variables from its original feature dimension to 10, and then back to its original dimension;  the first is followed by a batch normalization layer and ReLU activation, while the second is followed by ReLU activation only. 

\noindent\textbf{Training details }The adaptation training for our TARD module and the baseline methods was conducted with a consistent set of hyperparameters. We used the Adam optimizer with a learning rate of 1e-5. All models were trained with a batch size of 128. The basedline autoencoder was trained for 500 epochs. For TARD, the adaptive module ( $h_\phi$) was trained for 50 epochs, as its small size led to rapid convergence. During this phase, all batch normalization layers in the main autoencoder ($f_\theta$) were frozen. The AdaBN layer within our adaptive module, as well as in the AdaBN baseline, updates its running statistics on a per-batch basis during both adaptation training and subsequent testing.

\noindent\textbf{Baselines}
We evaluate the performance of our proposed method against two established domain adaptation techniques: Adaptive Batch Normalization (AdaBN) \cite{li2020adabn} and Maximum Mean Discrepancy (MMD) \cite{long2015dan}. These baselines were specifically selected to ensure a fair and relevant comparison under the strict constraints of our problem setting: complete absence of labeled fault data, limited normal target data, and no access to source data during online adaptation.
Both methods are unsupervised and do not require labels in the source or target domains, making them suitable for comparison with our approach. 
Furthermore, both they represent widely-adopted approaches for unsupervised domain adaptation—AdaBN for feature-level normalization strategies and MMD for explicit distribution alignment—making them appropriate benchmarks for evaluating our proposed method's. 
For MMD, we align the feature distributions of the source and target domains by minimizing the discrepancy during training, using the same target domain data as in our proposed approach. 
Methods such as Domain-Adversarial Neural Networks (DANN)~\cite{ganin2016domain}, which require labeled source data during adaptaion, are not included in our evaluation, as they operate under different assumptions and are therefore not directly comparable. Similarly, we exclude methods that require simultaneous access to source and target data during training due to practical deployment constraints where source data may be unavailable. 
Additionally, we include a baseline model trained exclusively on source domain data without any adaptation, serving as a reference to quantify the improvement achieved by domain adaptation strategies. 

\noindent\textbf{Evaluation metrics}
Model performance is evaluated using three standard metrics: accuracy, F1 score, and area under the ROC curve (AUC). Accuracy measures the proportion of correct predictions among all samples, offering a general sense of performance. However, in cases of class imbalance, in our case the imbalance between normal and faulty samples, it may not fully reflect model effectiveness. In addition, we evaluate the F1 score, which combines precision and recall into a single metric by computing their harmonic mean, providing a balanced view of the model's ability to correctly identify positive cases while minimizing false positives and false negatives. AUC measures the area under the Receiver Operating Characteristic curve, which plots the true positive rate against the false positive rate at various thresholds. It reflects the model's ability to distinguish between classes across different decision boundaries and is particularly useful for assessing classifier performance in imbalanced or uncertain environments.





\subsection{Case study 1: Cranfield three-phase flow facility}

\noindent\textbf{Dataset} The Three-Phase Flow Facility at Cranfield University is a laboratory-scale setup designed to provide a controlled and measurable flow of water, oil, and air within a pressurized system~\cite{ruiz2015statistical}. 
The facility consists of key industrial components, including multiple pipelines, a 10.5-meter-high riser, a two-phase separator, and a larger three-phase separator at ground level, along with compressors, pumps, and storage tanks. The entire process is managed by a SCADA system. The system is monitored by 24 different process variables, which include pressures, flow rates, levels, densities, temperatures, and valve positions. Table~\ref{tab:cranfield_variables} shows the monitored process variables. The two primary inputs used to control the operating conditions are the air and water flow rate set points. For model training, three datasets (T1, T2, and T3) were collected under normal operating conditions, with varying air and water flow rates to ensure data diversity.
In addition to the normal operation datasets, data were collected under faulty conditions, where six distinct faults were intentionally introduced to the system. These include air and water line blockages, a top separator input blockage, an open direct bypass to simulate a leakage, slugging conditions to create an unstable flow regime, and the pressurization of the 2" line to simulate an incorrect operational procedure. These faults were selected to simulate real-world malfunctions. The faults were introduced after a period of stable normal operation for the investigation of changes in health indicators. Whenever feasible, the faults were applied gradually, allowing observation of how increasing fault severity impacted the system behavior. Once the fault reached a predefined severity level, it was removed, and the system was returned to normal operating conditions.

Table~\ref{tab:cranfield_dataset} summarizes the fault cases introduced in the Three-Phase Flow Facility, providing details on the operation condition, time span, the ratio of faulty samples among all data samples. In this study, all the data was captured at a sampling rate of 1 Hz. Six fault types were included: air line blockage, water line blockage, top separator input blockage, open direct bypass, slugging conditions, and pressurization of the 2” line. In all cases, faults were introduced within defined time windows while the system operated under either steady or changing conditions. The duration and timing of each fault event varied across the datasets. For the slugging conditions, faults were injected multiple times within a single sub-dataset.

\begin{figure}
    \centering
    \includegraphics[width=0.9\linewidth]{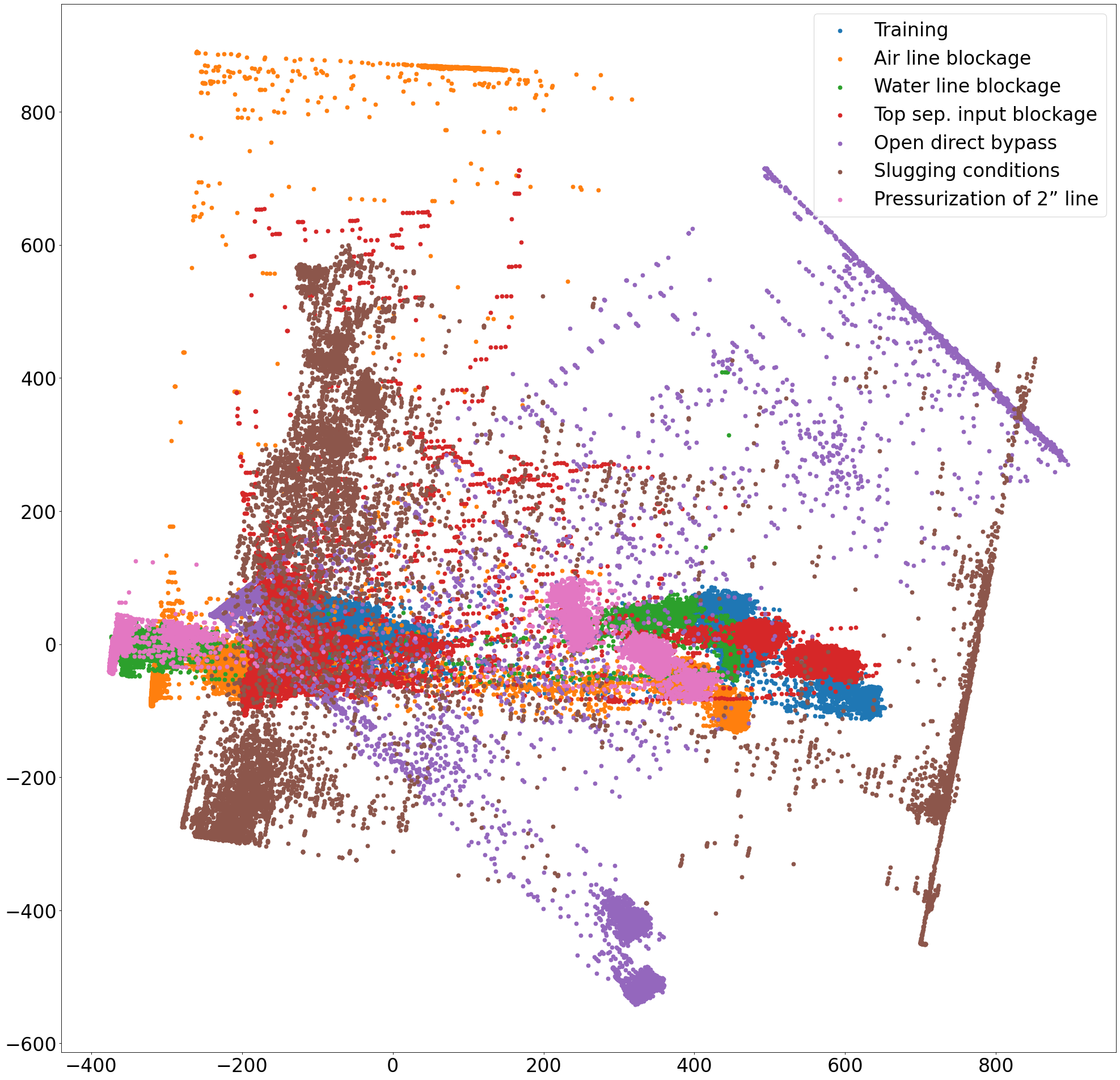}
    \caption{Visualization of the first two principle components of PCA analysis of the Cranfield three-phase flow facility dataset, including the normal training dataset and the testing dataset for each faulty condition.}
    \label{fig:cranfield_pca}
\end{figure}

We visualize the first two principal components of PCA of the Cranfield dataset in Figure~\ref{fig:cranfield_pca}, which capture 96$\%$ of the data's variability.
The visualization clearly confirms the domain shift between the broadly distributed source 'Training' data and the more localized target domains, which is the core challenge our method is designed to address. Furthermore, the plot highlights the distinct characteristics of the faults: some, like 'Slugging conditions,' form unique clusters indicating a strong signature, while the various blockage faults cluster more closely to the normal data, suggesting a more subtle detection challenge.

\renewcommand{\arraystretch}{0.8}
\begin{table}[H]
        \centering
\resizebox{1.0\textwidth}{!}{%
    \centering
    \begin{tabular}{c|c|l|c}

\hline
\textbf{Location} & \textbf{Measured Magnitude} & \textbf{Unit} \\
\hline
\multirow{2}{*}{Control variables}
& FT305  & Flow rate input air                            & Sm\textsuperscript{3}/s \\
& FT104  & Flow rate input water                          & kg/s    \\
\hline
\multirow{24}{*}{Measurement variables}
&PT312  & Air delivery pressure                          & MPa     \\
& PT401  & Pressure in the bottom of the riser            & MPa     \\
& PT408  & Pressure in top of the riser                   & MPa     \\
& PT403  & Pressure in top separator                      & MPa     \\
& PT501  & Pressure in 3 phase separator                  & MPa     \\
& PT408  & Diff. pressure (PT401-PT408)                   & MPa     \\
& PT403  & Differential pressure over VC404               & MPa     \\
& FT407  & Flow rate top riser                            & kg/s    \\
& LI405  & Level top separator                            & m       \\
& FT406  & Flow rate top separator output                 & kg/s    \\
& FT407  & Density top riser                              & kg/m\textsuperscript{3} \\
& FT406  & Density top separator output                   & kg/m\textsuperscript{3} \\
& FT104  & Density water input                            & kg/m\textsuperscript{3} \\
& FT407  & Temperature top riser                          & \textdegree C \\
& FT406  & Temperature top separator output               & \textdegree C \\
& FT104  & Temperature water input                        & \textdegree C \\
& LI504  & Level gas-liquid 3 phase separator             & \%      \\
& VC501  & Position of valve VC501                        & \%      \\
& VC302  & Position of valve VC302                        & \%      \\
& VC101  & Position of valve VC101                        & \%      \\
& PO1    & Water pump current                             & A       \\
& PT417  & Pressure in mixture zone 2'' line              & MPa     \\
        \hline
    \end{tabular}
    \caption{Cranfield three-phase flow facility dataset variables}
    \label{tab:cranfield_variables}
    }
\end{table}

\renewcommand{\arraystretch}{0.8}
\begin{table}[H]
\resizebox{1.0\textwidth}{!}{%
    \centering
    \begin{tabular}{l|c|c|c}
        \hline
        \textbf{Fault case} & \textbf{Operation condition} & \textbf{Duration (s)}& Fault ratio \\
        \hline
        \multirow{3}{*}{Air line blockage} & changing & 5811 & 0.62 \\
        & steady & 4467 & 0.70 \\
        & steady & 4321 & 0.69 \\
        \hline
        \multirow{3}{*}{Water line blockage} 
        & changing & 9192  & 0.48 \\
        & steady  & 3496  & 0.62 \\
        & steady & 3421  & 0.62 \\
        \hline
        \multirow{3}{*}{Top separator input blockage}
        & changing & 9090 & 0.79 \\
        & steady & 6272 & 0.88 \\
        & steady & 10764 & 0.83 \\
        \hline
        \multirow{3}{*}{Open direct bypass}
        & changing & 7208 & 0.74 \\
        & steady & 4451 & 0.67 \\
        & steady & 3661 & 0.82 \\
        \hline
        \multirow{5}{*}{Slugging conditions} 
        & \multirow{2}{*}{changing} & \multirow{2}{*}{2541}
        & 0.43 \\
        & & &0.19 \\
        \cline{2-4}
        & \multirow{3}{*}{changing} & \multirow{3}{*}{10608}
        & 0.12 \\
        & & & 0.05 \\
        & & & 0.24 \\
        \hline
        \multirow{2}{*}{Pressurization of the 2” line}
        & changing & 2800 & 0.38 \\
        & changing & 4830 & 0.79 \\
        \hline
    \end{tabular}
    \caption{Summary of introduced faults  of cranfield three-phase flow facility dataset}
    \label{tab:cranfield_dataset}
}
\end{table}

\noindent\textbf{Implementation details} 
Our objective is to enable robust fault detection with limited data availability. A source model is first pretrained with abundant source data, and then adapted to target operation conditions that have limited collected data samples. 
In the Cranfield case study, the training dataset (T1, T2 and T3) collected under varying normal operating conditions is selected as the source domain. Each faulty case study is treated as a separate target domain, where only a limited number of training samples are available. Specifically, for each fault case, we use 2500 normal samples for training and evaluate the model on the remaining data.
During testing, the training data from each target domain is utilized to compute the anomaly detection threshold $\bar{r}_{t_training}$. The detection sensitivity parameter is set to $\alpha = 1.0$ for all case studies in this experiment.

\noindent\textbf{Fault detection result}
Table~\ref{tab:result_cran} summarizes the fault detection performance across six fault cases, comparing our proposed TARD method with the baseline, AdaBN, and MMD adaptation methods. 
Overall, the proposed TARD method consistently outperforms the other methods across different faulty cases, highlighting its effectiveness and robustness in fault detection,  particularly under limited training data conditions.
While the baseline and other adaptation methods perform relatively poorly in challenging cases such as airline blockage and open direct bypass (F1 scores around or below 0.56), TARD achieves notable improvements, reaching F1 scores of 0.70 and 0.69, respectively. These results indicate that TARD is more robust in detecting subtle or less distinguishable fault patterns.
In fault cases which are relatively easy to detect, such as slugging conditions and pressurization of the 2" line, most methods already achieve high performance, while TARD still maintains the highest or comparable metrics, with F1 scores of 0.92 and 1.00, demonstrating consistent improvement across various adaptation scenarios.

The superior performance of TARD can be attributed to its dedicated adaptation strategy. Unlike AdaBN, which adjusts all model parameters and can overfit to batch statistics of potential faulty samples, and MMD, which aligns global distributions and may miss local shifts online, TARD's separate adaptive module focused on control variables allows it to precisely compensate for operational changes without distorting the learned representation of normal sensor behavior. This targeted adaptation is particularly effective in reducing false alarms caused by distribution shifts.

\renewcommand{\arraystretch}{1.0}
\begin{table}[h]
\centering
\resizebox{1.0\textwidth}{!}{%
\begin{tabular}{l|ccc|ccc|ccc|ccc}
\toprule
\multirow{2}{*}{\textbf{Fault Case}} 
& \multicolumn{3}{c|}{\textbf{Baseline}} 
& \multicolumn{3}{c|}{\textbf{AdaBN}} 
& \multicolumn{3}{c|}{\textbf{MMD}} 
& \multicolumn{3}{c}{\textbf{TARD (ours)}} \\
& \textbf{Acc} & \textbf{F1} & \textbf{AUC} 
& \textbf{Acc} & \textbf{F1} & \textbf{AUC} 
& \textbf{Acc} & \textbf{F1} & \textbf{AUC} 
& \textbf{Acc} & \textbf{F1} & \textbf{AUC}  \\
\midrule
Air line blockage          & 0.50 & 0.43 & 0.61 & 0.52 & 0.43 & 0.61 & 0.53 & 0.47 & 0.64 & 0.62 & 0.70 & 0.61 \\
Water line blockage        & 0.63 & 0.67 & 0.63 & 0.60 & 0.62 & 0.60 & 0.66 & 0.69 & 0.66 & 0.68 & 0.76 & 0.67 \\
Top sep. input blockage    & 0.55 & 0.63 & 0.73 & 0.57 & 0.65 & 0.74 & 0.56 & 0.64 & 0.73 & 0.72 & 0.79 & 0.83 \\
Open direct bypass         & 0.51 & 0.53 & 0.65 & 0.56 & 0.60 & 0.66 & 0.53 & 0.56 & 0.66 & 0.62 & 0.69 & 0.67 \\
Slugging conditions        & 0.83 & 0.85 & 0.86 & 0.88 & 0.88 & 0.89 & 0.90 & 0.89 & 0.90 & 0.93 & 0.92 & 0.93 \\
Pressurization of 2" line  & 0.93 & 0.94 & 0.95 & 0.98 & 0.98 & 0.97 & 1.00 & 1.00 & 1.00 & 1.00 & 1.00 & 1.00 \\
\bottomrule
\end{tabular}
}
\caption{Cranfield three-phase flow facility dataset fault detection results for different domain adaptation methods. Metrics shown are Accuracy, F1-score, and AUC.}
\label{tab:result_cran}
\end{table}

\begin{figure}[H]
    \centering
    \includegraphics[width=1.0\linewidth]{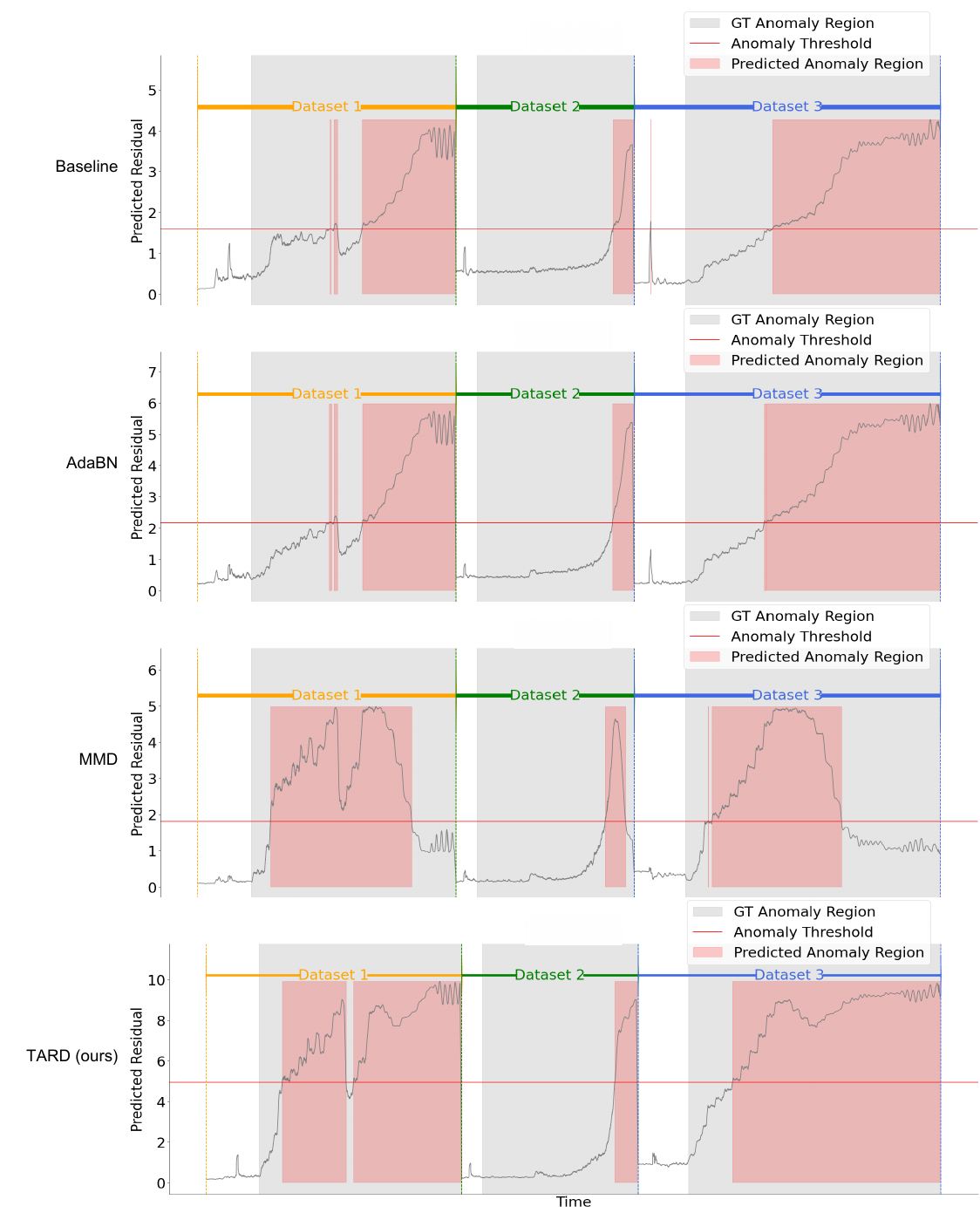}
    \caption{Anomaly detection residuals for the Cranfield Top separator input blockage case. The gray regions indicate the reported fault periods for each dataset subset and the red regions indicate the predicted anomaly region. The red line denotes the computed anomaly detection threshold $\alpha *\bar{r}_{t\_training}$.}
    \label{fig:case3_cranfield}
\end{figure}

To further investigate the detection performance, we present a visualization of the top separator input blockage fault case, illustrating the reconstruction residuals alongside the anomaly detection thresholds in Figure~\ref{fig:case3_cranfield}. As shown in the figure, both the baseline and AdaBN methods exhibit difficulty in adapting to the evolving operating conditions, resulting in delayed anomaly detection and missed early fault indications. The MMD method shows signs of overfitting to the limited target domain samples during the adaptation process, leading to false negatives and reduced robustness. In contrast, our proposed method effectively adapts to the gradual operational changes from an early stage, maintaining stable residuals and enabling early and accurate detection of anomalies. This case study shows the robustness of the proposed approach in handling complex and gradually developing fault patterns, where both insufficient adaptation and overfitting can compromise detection reliability.

A key observation in Figure~\ref{fig:case3_cranfield} is the apparent delay between the start of the reported ground truth (GT) fault period and the anomaly detected by TARD. This is an expected result that reflects the physical nature of the fault. The 'Top separator input blockage' is a gradually developing fault. The GT period marks the initiation of the blockage, but it takes time for its effects to propagate and manifest as a significant deviation in the sensor readings. The initial, subtle changes are characteristic of an incipient fault, which may not be distinguishable from normal process noise. Our method, with its threshold set for robustness, correctly identifies the fault once its signature becomes statistically significant and persistent, thus ensuring reliable detection while avoiding false alarms from minor, ambiguous fluctuations.

\noindent\textbf{Ablation study on Cranfield Top sep. input blockage case}

We present the fault detection results of Cranfield Top sep. input blockage case under different operating conditions in Table~\ref{tab:result_cran_topsep}. As observed, TARD achieves comparable or superior performance in both steady and changing conditions. In contrast, MMD tends to overfit to steady-state data, which diminishes its fault detection capability under varying conditions. TARD, on the other hand, demonstrates a consistently stable performance. Notably, TARD brings the most significant improvements under changing operating conditions — a particularly challenging scenario that remains insufficiently addressed by existing unsupervised adaptation methods such as MMD. Although AdaBN is a test-time adaptation method, it fails to yield performance improvements in this case study due to its sensitivity and inadaptability to abrupt changes in dynamic environments with rapid changes.

\begin{table}[h]
\centering
\resizebox{1.0\textwidth}{!}{%
\begin{tabular}{l|ccc|ccc|ccc|ccc}
\toprule
\multirow{2}{*}{\textbf{Operating conditions}} 
& \multicolumn{3}{c|}{\textbf{Baseline}} 
& \multicolumn{3}{c|}{\textbf{AdaBN}} 
& \multicolumn{3}{c|}{\textbf{MMD}} 
& \multicolumn{3}{c}{\textbf{TARD (ours)}} \\
& \textbf{Acc} & \textbf{F1} & \textbf{AUC} 
& \textbf{Acc} & \textbf{F1} & \textbf{AUC} 
& \textbf{Acc} & \textbf{F1} & \textbf{AUC} 
& \textbf{Acc} & \textbf{F1} & \textbf{AUC}  \\
\midrule
Changing & 0.58&0.64&0.74 &0.58&0.64&0.74 &0.73&0.79&0.83 &0.81&0.86&0.88 \\
Steady & 0.28&0.32&0.56 &0.30&0.35&0.61 &0.34&0.41&0.63 &0.30&0.34&0.60\\
Steady &0.63&0.71&0.78 &0.66&0.74&0.79 &0.56&0.64&0.73 &0.80&0.86&0.89\\
\bottomrule
\end{tabular}
}
\caption{Cranfield Top sep. input blockage case fault detection results for different domain adaptation methods. Metrics shown are Accuracy, F1-score, and AUC.}
\label{tab:result_cran_topsep}
\end{table}

To further discuss the uncertainty in our fault detection results, we conducted an uncertainty quantification (UQ) experiment for this Cranfield Top separator input blockage case study, where we implemented a Deep Ensemble for estimating the uncertainty in the anomaly score.
We trained 10 independent TARD models with identical architectures but different random weight initializations. During inference, each test batch is processed by all 10 models in the ensemble. This yields a distribution of 10 different anomaly scores for each data point. The mean of this distribution is used as the final, more robust anomaly score, while the standard deviation serves as a direct measure of the models' disagreement, or uncertainty.

\begin{figure}[H]
    \centering
    \includegraphics[width=1.0\linewidth]{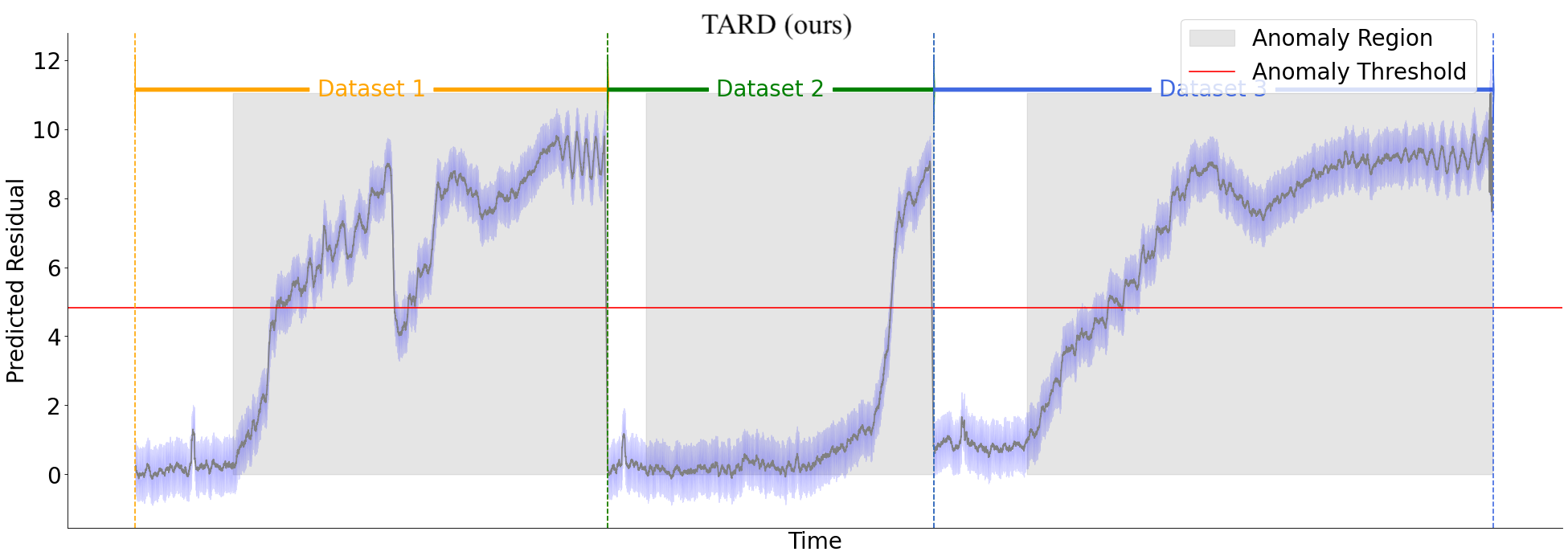}
    \caption{Uncertainty estimation for the Cranfield Top separator input blockage case. The gray regions indicate the reported fault periods for each dataset subset and the red regions indicate the predicted anomaly region. The red line denotes the computed anomaly detection threshold $\alpha *\bar{r}_{t\_training}$. The shaded blue confidence interval represents mean anomaly score ± one standard deviation}
    \label{fig:confidence}
\end{figure}

Figure~\ref{fig:confidence} displays the mean anomaly score from the deep ensemble, with the shaded blue band representing the model uncertainty (± one standard deviation).
During the detected fault period, the uncertainty band remains consistently narrow, with the standard deviation averaging approximately 0.8. This low level of disagreement among the ensemble models signifies a high-confidence consensus. This result confirms that the detection is not only effective but also highly robust.

\subsection{Case study 2: PRONTO heterogeneous benchmark}

\noindent\textbf{Dataset}
The Pronto heterogeneous benchmark dataset is an industrial-scale high-pressure multiphase flow facility for the investigation of the dynamics and control of multiphase flows. The process for this study involved mixing water and air, which then flowed through a horizontal pipeline to a 2-inch riser with an S-shaped section before being separated; the oil phase was not used. The facility is heavily instrumented with standard pressure, flow, temperature, and density sensors, as well as high-frequency pressure and ultrasonic sensors. Table~\ref{tab:pronto_variables} shows the monitored process variables. 
A total of 20 distinct operating conditions are covered in the dataset, characterized by various combinations of air and water flow rates. 
Table~\ref{tab:pronto_dataset}. summarizes the fault cases introduced in the dataset, providing details on the operation condition, time span, the ratio of faulty samples among all data samples. In this study, all the data was captured at a sampling rate of 1 Hz.
The dataset incorporates three types of manually induced faults, including air leakage, air blockage, and diverted flow.
Air leakage involves a gradual loss of air supply, leading to cyclic flow behavior and eventually a transition to a liquid-only regime. Air blockage restricts the air supply without reducing system pressure, resulting in milder forms of slugging. Diverted flow reroutes the flow, altering the regime, particularly at the riser section. Each fault type is applied at two different operating points, with the severity adjusted incrementally. 
Each of these faults was introduced at two different operating points by incrementally adjusting valve openings, allowing the fault severity to be progressively varied. The dataset includes precise timestamps, process variable measurements, and supplementary data such as alarm logs, high-frequency sensor readings, and visual recordings, all aligned with the induced faults. For comprehensive fault descriptions, please refer to~\cite{stief2019heterogeneous}.

\renewcommand{\arraystretch}{0.8}
\begin{table}[H]
\resizebox{0.6\textwidth}{!}{%
    \centering
    \begin{tabular}{l|c|c}
        \hline
        \textbf{Fault case} & \textbf{Duration (s)}& Fault ratio \\
        \hline
        Air leakage & 6009 & 0.21 \\
        \hline
        Air blockage & 5947 & 0.20 \\
        \hline
        Diverted flow & 7143 & 0.33 \\
        \hline
        
    \end{tabular}
    \caption{Summary of introduced faults  of PRONTO dataset}
    \label{tab:pronto_dataset}
}
\end{table}

We visualize the first two principal components of PCA of the PRONTO dataset, which account for $92\%$ of the data's variability.
The plot clearly illustrates the domain shift between the 'Normal' source data (blue) and the three fault conditions. The separation confirms that the operating points where faults were introduced are distinct from the source domain, validating the need for an adaptation method like TARD. Furthermore, the visualization reveals the complex, non-linear structure of the fault data. While 'Air blockage' (green) forms relatively tight clusters, 'Air leakage' (orange) and especially 'Diverted flow' (red) are widely distributed, highlighting the challenging nature of detecting these conditions. The significant overlap between fault types underscores the focus of our work on fault detection (distinguishing normal from any fault) in complex systems.

\begin{figure}
    \centering
    \includegraphics[width=0.9\linewidth]{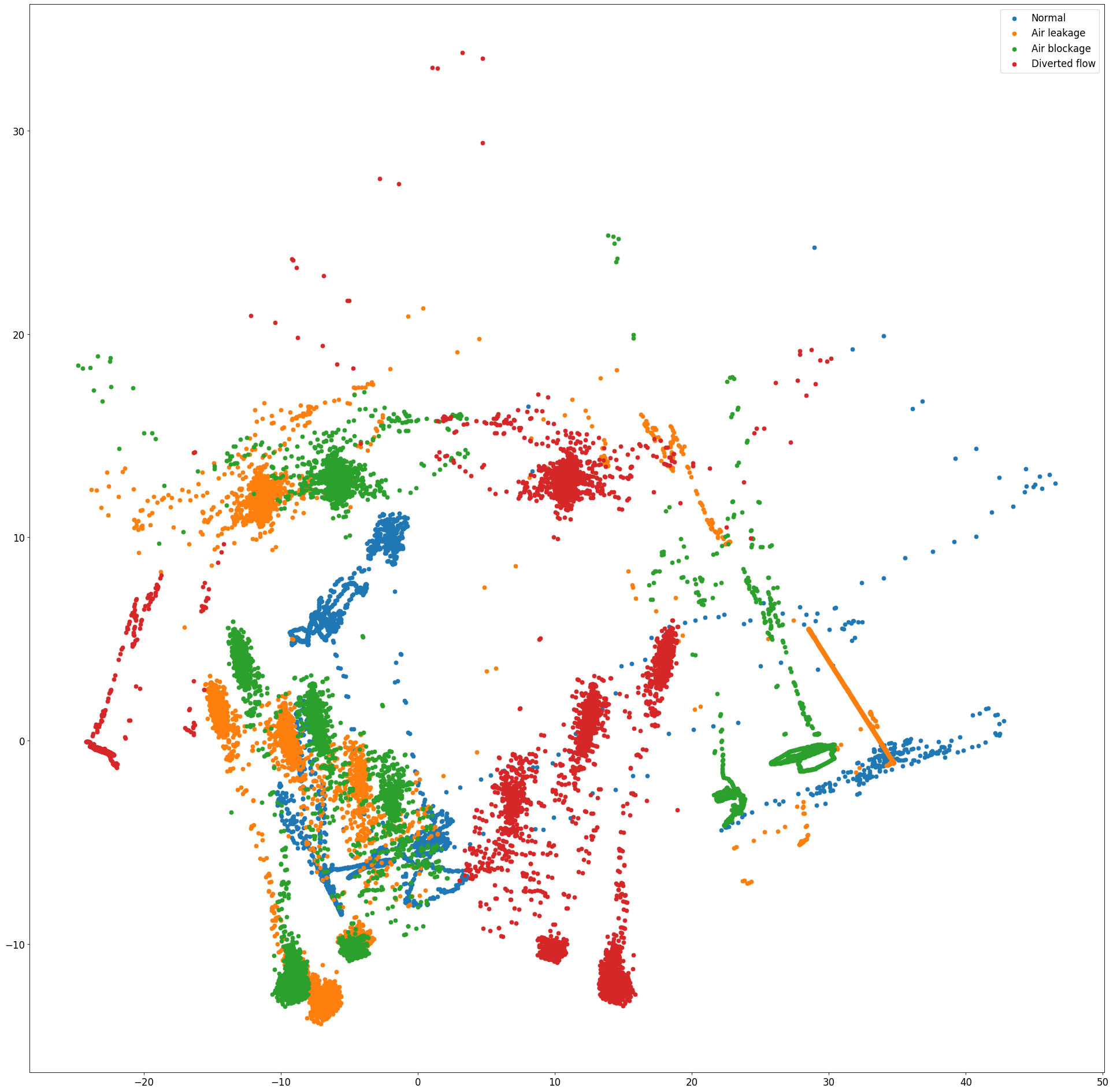}
    \caption{Visualization of the first two principle components of PCA analysis of the PRONTO heterogeneous dataset, including the normal training dataset and the testing dataset for each faulty condition.}
    \label{fig:pronto_pca}
\end{figure}





\renewcommand{\arraystretch}{0.8}
\begin{table}[H]
        \centering
\resizebox{1.0\textwidth}{!}{%
    \centering
    \begin{tabular}{c|c|l|c}

\hline
\textbf{Location} & \textbf{Measured Magnitude} & \textbf{Unit} \\
\hline
\multirow{4}{*}{Control variables}
& FT305/302 & Input air flow rate                          & Sm\textsuperscript{3} h\textsuperscript{-1} \\
& FT102/104 & Input water flow rate                        & kg s\textsuperscript{-1} \\
& FT305-T   & Input air temperature                        & \textdegree C \\
& FT102-T   & Input water temperature                      & \textdegree C \\
\hline
\multirow{13}{*}{Measurement variables}

& PT312     & Air delivery pressure                        & bar(g) \\
& FT102-D   & Input water density                          & kg m\textsuperscript{-3} \\
& PT417     & Pressure in the mixing zone                  & bar(g) \\
& PT408     & Pressure at the riser top                    & bar(g) \\
& PT403     & Pressure in the 2-phase separator            & bar(g) \\
& FT404     & 2-phase separator output air flow rate       & m\textsuperscript{3} h\textsuperscript{-1} \\
& FT406     & 2-phase separator output water flow rate     & kg s\textsuperscript{-1} \\
& PT501     & Pressure in the 3-phase separator            & bar(g) \\
& PIC501    & Air outlet valve 3-phase separator           & \% \\
& LI502     & Water level 3-phase separator                & \% \\
& LI503     & Water coalescer level                        & \% \\
& LVC502    & Water coalescer outlet valve                 & \% \\
& LI101     & Water tank level                             & m \\
        \hline
    \end{tabular}
    \caption{PRONTO heterogeneous benchmark dataset variables}
    \label{tab:pronto_variables}
    }
\end{table}

\noindent\textbf{Implementation details} 
In the PRONTO case study, the initial 3000 samples collected under normal operating conditions are used as the source domain for pretraining. Additionally, 500 samples from the target domain are used for adaptation training. For each fault case, the model is evaluated using the remaining normal operating data combined with the corresponding faulty data to perform fault detection testing.
During the testing phase, the adaptation training data is also used to compute the anomaly detection threshold $\bar{r}_{t_{training}}$. The detection sensitivity parameter is set to $\alpha = 1.0$ for all case studies in this experiment.

\noindent\textbf{Fault detection result}
Table~\ref{tab:result_pronto} presents the fault detection performance on the PRONTO heterogeneous benchmark dataset. Across all three fault cases—air leakage, air blockage, and diverted flow—TARD consistently achieves the highest F1-scores and AUC values, demonstrating its robustness and generalization across heterogeneous domains.

For air leakage, a relatively subtle fault, TARD achieves an F1 score of 0.76, significantly outperforming the baseline (0.62), AdaBN (0.36), and MMD (0.51). In the air blockage case, which is less challenging given its more obvious faulty signal pattern, all methods perform well, while TARD maintains competitive performance, reaching an F1 score of 0.94 and an AUC of 0.98, slightly improving upon other methods.

For the diverted flow scenario, which poses significant challenges due to its subtle and variable patterns, TARD achieves an F1 score of 0.69 and an AUC of 0.78, markedly outperforming the baseline (F1 0.11, AUC 0.50) and the adaptation methods AdaBN and MMD (both F1 0.51, AUC 0.62 and 0.60, respectively). These results highlight TARD's ability to handle difficult fault patterns where conventional approaches struggle to maintain reliable detection performance.

\begin{table}[h]
\centering
\resizebox{1.0\textwidth}{!}{%
\begin{tabular}{l|ccc|ccc|ccc|ccc}
\toprule
\multirow{2}{*}{\textbf{Fault Case}} 
& \multicolumn{3}{c|}{\textbf{Baseline}} 
& \multicolumn{3}{c|}{\textbf{AdaBN}} 
& \multicolumn{3}{c|}{\textbf{MMD}} 
& \multicolumn{3}{c}{\textbf{TARD (ours)}} \\
& \textbf{Acc} & \textbf{F1} & \textbf{AUC} 
& \textbf{Acc} & \textbf{F1} & \textbf{AUC} 
& \textbf{Acc} & \textbf{F1} & \textbf{AUC} 
& \textbf{Acc} & \textbf{F1} & \textbf{AUC}  \\
\midrule
Air leakage &0.81&0.62&0.78 &0.81&0.36&0.63 &0.74&0.51&0.74 &0.87&0.76&0.92 \\
Air blockage &0.97&0.93&0.96 &0.94&0.88&0.96 &0.99&0.96&0.97 &0.98&0.94&0.98 \\
Diverted flow &0.67&0.11&0.50 &0.66&0.51&0.62 &0.65&0.51&0.60 &0.72&0.69&0.78 \\
\bottomrule
\end{tabular}
}
\caption{PRONTO heterogeneous benchmark dataset fault detection results for different domain adaptation methods. Metrics shown are Accuracy, F1-score, and AUC.}
\label{tab:result_pronto}
\end{table}

The results on the PRONTO dataset further validate TARD's effectiveness. The significant improvement on the "diverted flow" fault, which is particularly subtle, underscores the advantage of TARD's targeted adaptation. By focusing adaptation on the control variables, TARD avoids overfitting to the overall data distribution and maintains sensitivity to nuanced fault signatures that other methods miss.

\begin{table}[h]
\centering
\resizebox{1.0\textwidth}{!}{%
\begin{tabular}{l|ccc|ccc|ccc|ccc}
\toprule
\multirow{2}{*}{\textbf{Fault Case}} 
& \multicolumn{3}{c|}{\textbf{Baseline}} 
& \multicolumn{3}{c|}{\textbf{AdaBN}} 
& \multicolumn{3}{c|}{\textbf{MMD}} 
& \multicolumn{3}{c}{\textbf{TARD (ours)}} \\
& \textbf{Acc} & \textbf{F1} & \textbf{AUC} 
& \textbf{Acc} & \textbf{F1} & \textbf{AUC} 
& \textbf{Acc} & \textbf{F1} & \textbf{AUC} 
& \textbf{Acc} & \textbf{F1} & \textbf{AUC}  \\
\midrule
Air leakage &0.81&0.62&0.78 &0.81&0.36&0.63 &0.74&0.51&0.74 &0.87&0.76&0.92 \\
Air blockage &0.97&0.93&0.96 &0.94&0.88&0.96 &0.99&0.96&0.97 &0.98&0.94&0.98 \\
Diverted flow &0.67&0.11&0.50 &0.66&0.51&0.62 &0.65&0.51&0.60 &0.72&0.69&0.78 \\
\bottomrule
\end{tabular}
}
\caption{PRONTO heterogeneous benchmark dataset fault detection results for different domain adaptation methods. Metrics shown are Accuracy, F1-score, and AUC.}
\label{tab:result_pronto}
\end{table}

\subsection{Case study 3: High-dimensional synthetic dataset}

To validate the scalability and performance in high-dimensional settings, we generated a synthetic benchmark dataset.
The simulated features include control variables $w$, sensor measurements $x$, and normalized time index $t \in [0, 1]$ to simulate a gradual temporal shift. The input features consist of independent control variables $w$. A high-dimensional set of sensor measurements $x$ was generated as non-linear functions of these control variables. To simulate a realistic industrial environment, the sensor responses were designed with intricate interdependencies, including:
\begin{itemize}
    \item Temporal Domain Drift: We incorporate the evolution over time by linearly drifting some of the sensor measurements linearly over time. This simulates real-world phenomena such as equipment wear and tear, gradual calibration shifts, or changing environmental conditions, ensuring the "normal" state is a moving target.
    \item Redundancy and Correlation: To simulate the interdependency common in industrial systems, a subset of sensors was defined as linear combinations of others. These redundant sensors inherently inherit the drift of their source signals. 
    \item Stochastic Noise: Gaussian noise was added to each sensor signal to mimic real-world measurement imprecision.
\end{itemize}
We generate datasets for two sets of experiments: one with $100$ sensor measurements and $10$ control variables, and another in a higher-dimensional setting with $1000$ sensor measurements and $100$ control variables. For each setting, we generate normal data of 2{,}000 timesteps for pre-training. To simulate abnormal behavior, we synthesize a test set of 5{,}000 data samples and insert two contiguous anomaly periods covering approximately 10\% of the 5{,}000 timesteps. During each anomaly window, a random subset of state features is shifted by a constant offset. 
All timesteps within these windows are labeled anomalous ($\texttt{label}=1$), while the rest are normal ($\texttt{label}=0$). The final dataset therefore has shape $5000 \times 111$, consisting of 
100 state features, 10 control features, and 1 binary label column. 

Table~\ref{tab:high-dimensi} summarizes the detection results and computational requirements for both scenarios.In both settings, TARD significantly outperforms the baseline. Although the performance is challenged in the extreme case of 1000 sensor measurements, it still provides a substantial improvement over the baseline, which struggles significantly in this high-dimensional space. The results also demonstrate that TARD's adaptation process is highly efficient. The inference latency remains well within the requirements for real-time monitoring, confirming that TARD efficiently scales for high-dimensional data.

\renewcommand{\arraystretch}{2.0}
\begin{table}[H]
\resizebox{1.0\textwidth}{!}{%
    \centering
       \begin{tabular}{|c|c|c|c|c|c|c|}
\hline
 \multirow{2}{*}{\textbf{\makecell{Sensor \\ Measurements}}} & \multirow{2}{*}{\textbf{Method}} & \multicolumn{2}{c|}{\textbf{Detection Result}} & \multicolumn{3}{c|}{\textbf{Computational Efficiency}} \\
\cline{3-7}
 & &  \textbf{Acc} & \textbf{F1} & \textbf{Training Time (s)} & \textbf{Inference Time (ms)} & \textbf{\makecell{Number of \\ Model Parameters}} \\
\hline
\multirow{2}{*}{100}  & baseline & 0.80 & 0.42 & 47.83 & 0.51$\pm$0.08& 46100 \\
\cline{2-7}
& TARD & 0.89 & 0.67 & 16.04 & 0.97$\pm$0.07& 93550 \\
\hline
\multirow{2}{*}{1000}  & baseline & 0.73 & 0.10 & 54.53 & 0.75$\pm$0.08& 1226900 \\
\cline{2-7}
& TARD & 0.76 & 0.48 & 16.50 & 1.60$\pm$0.02& 4442350 \\

\hline
\end{tabular}
    \caption{Experimental results on high-dimensional synthetic dataset.}
    \label{tab:high-dimensi}
}
\end{table}

The TARD framework is well-suited for systems with many sensors and complex dependencies, as confirmed by these results. The core of the model is a reconstruction-based autoencoder that is pre-trained offline on the source dataset. This autoencoder, composed of fully-connected layers, is chosen to effectively learn a low-dimensional representation of the non-linear relationships hidden within the high-dimensional sensor data. Crucially, the online adaptation is handled by a separate, lightweight adaptive module. This module is intentionally kept simple, consisting of only two fully connected layers. Its size and complexity scale only with the number of control variables, not the total number of sensors, as it exclusively processes control variables as its input.

The computational and memory loads are intentionally split between offline and online phases to ensure real-time feasibility. The primary memory requirement is for storing the weights of the pre-trained autoencoder, which is a one-time cost determined before deployment. The most computationally intensive phase is the initial training of the autoencoder on the source data, but since this is performed only once, it does not impact the model's operational performance. At test time, the process is 
highly efficient. It involves a single forward pass through the frozen autoencoder  and the very small adaptive module. The adaptation is continuous during inference, with an AdaBN layer in the adaptive module updating its statistics based on each test batch. This lean online process makes TARD suitable for real-time fault detection. 

\color{black}

\section{Conclusions}

In this paper, we propose TARD, an effective continuous test-time domain adaptation approach for efficient and robust fault detection under evolving operating conditions in complex industrial systems. Unlike conventional fault detection models that rely heavily on large-scale labeled datasets or assume static environments, TARD is able to adapt to novel domain shifts in real time under real-world constraints—specifically, the absence of labeled faulty data and the limited availability of normal samples. This makes TARD highly align with the practical needs of real-world industrial systems: the scarcity of comprehensive, labeled data and the prohibitive cost of frequent re-training under non-stationary system behavior.

TARD leverages a lightweight adaptation module that continuously dapts during inference, allowing it to respond to gradual or abrupt domain shifts. Through extensive experimentation in comparison with two other representative domain adaptation methods. we demonstrate that TARD not only achieves early fault detection but also maintains a low false alarm rate under significant domain shifts, including variations between different stations and temporal drifts in system behavior.


Despite its strong performance, we have identified key directions for future enhancement to bolster TARD’s autonomy and long-term reliability in industrial settings. A primary area for improvement is the development of a robust \textbf{safeguard mechanism} to handle major, permanent system changes. This would involve continuously \textbf{quantifying the domain shift} in control variables and using a threshold to \textbf{trigger a targeted retraining} of the adaptive module ($h_{\phi}$) when a significant shift is detected. This ensures the model remains synchronized with the system’s evolving baseline without requiring a full reset. Furthermore, developing a method for the \textbf{automatic adjustment} of the fault detection sensitivity parameter ($\alpha$) would create a more adaptive and self-sufficient system, optimizing the trade-off between minimizing false alarms and ensuring prompt fault detection in dynamic environments. Finally, the fault analysis can be enhanced by examining the temporal dynamics of the anomaly scores themselves. While our current approach smooths residuals to ensure robustness against transient sensor noise, this trades temporal detail for reliability. A powerful next step is to introduce a secondary analysis module, such as a dedicated time-series model, to specifically detect complex patterns like slow-growing trends or cyclical behaviors within the residual sequence.

In conclusion, TARD demonstrates that continuous test-time adaptation is not only feasible but also highly effective for real-world fault detection tasks under non-stationary conditions. By addressing key limitations in future work, TARD could evolve into a general-purpose solution for intelligent fault monitoring across a wide range of industrial systems.

\section*{Acknowledgment of AI Assistance in Manuscript Preparation}
During the preparation of this work, the authors used ChatGPT to assist with refining and correcting the text. After using this tool, the authors carefully reviewed and edited the content as needed and take full responsibility for the content of this publication.

\section*{Acknowledgment of project funding}
This research was funded by the Innosuisse-Swiss Innovation Agency under project 100.574 IP-ICT - Scalable pump fleet knowledge transfer (FLEET). We thank the Innosuisse-Swiss Innovation Agency for their support for this research and paper.



\color{black}
\bibliographystyle{unsrt}  
\bibliography{PHME_Latex_Template}  

\end{document}